\documentclass[conference]{IEEEtran}
\IEEEoverridecommandlockouts

\usepackage{amssymb, amsmath}
\usepackage{amsfonts}
\usepackage{bbm}
\usepackage{mathrsfs}
\usepackage{xspace}
\usepackage{bm}

\usepackage{cite}
\usepackage{epsfig}
\usepackage{cases}

\newtheorem{proposition}{Proposition}

\newenvironment{textbmatrix}{   \setlength{\arraycolsep}{2.5pt}%
                                                                \big[\begin{matrix}}{\end{matrix}\big]%
                                                                \raisebox{0.08ex}{\vphantom{M}}}

\def\be{\begin{equation}}
\def\ee{\end{equation}}
\def\een{\nonumber \end{equation}}
\def\mat{\begin{bmatrix}}
\def\emat{\end{bmatrix}}
\def\btm{\begin{textbmatrix}}
\def\etm{\end{textbmatrix}}

\def\ba#1\ea{\begin{align}#1\end{align}}
\def\bs#1\es{\begin{split}#1\end{split}} 
\def\bg#1\eg{\begin{gather}#1\end{gather}} 
\def\bi#1\ei{\begin{itemize}#1\end{itemize}}

\newcommand{\safemath}[2]{\newcommand{#1}{\ensuremath{#2}\xspace}}

\DeclareMathOperator{\diag}{diag}                       

\safemath{\interior}{\mathrm{Int}}                       
\newcommand{\tp}[1]{\ensuremath{#1^{T}}}                
\newcommand{\herm}[1]{\ensuremath{#1^{H}}}      

\safemath{\dfn}{:=}                                                     
\safemath{\dirac}{\delta}                                       

\safemath{\SNR}{\text{\sc snr}}                                 
\safemath{\No}{N_0}                                                     
\safemath{\Es}{E_s}                                                     
\safemath{\Eb}{E_b}                                                     
\safemath{\EbNo}{\frac{\Eb}{\No}}
\safemath{\EsNo}{\frac{\Es}{\No}}

\DeclareMathOperator{\CHop}{\ensuremath{\mathbb{H}}} 
\safemath{\tvir}{h_{\CHop}}                                     
\safemath{\tvtf}{L_{\CHop}}                                     
\safemath{\spf}{S_{\CHop}}                                              
\safemath{\bff}{H_{\CHop}}                                      

\safemath{\ircf}{R_{h}}                                         
\safemath{\scf}{R_{S}}                                          
\safemath{\tfcf}{R_{L}}                                         
\safemath{\bfcf}{R_{H}}                                         

\safemath{\mi}{I}                                                       
\safemath{\capacity}{C}                                         

\safemath{\normal}{\mathcal{N}}                         
\safemath{\circnorm}{\mathcal{CN}}                      
\safemath{\mchain}{\leftrightarrow}                     

\safemath{\dB}{\,\mathrm{dB}}
\safemath{\dBm}{\,\mathrm{dBm}}
\safemath{\Hz}{\,\mathrm{Hz}}
\safemath{\kHz}{\,\mathrm{kHz}}
\safemath{\MHz}{\,\mathrm{MHz}}
\safemath{\GHz}{\,\mathrm{GHz}}
\safemath{\s}{\,\mathrm{s}}
\safemath{\ms}{\,\mathrm{ms}}
\safemath{\mus}{\,\mathrm{\mu s}}
\safemath{\ns}{\,\mathrm{ns}}
\safemath{\meter}{\,\mathrm{m}}
\safemath{\mm}{\,\mathrm{mm}}
\safemath{\cm}{\,\mathrm{cm}}
\safemath{\m}{\,\mathrm{m}}
\safemath{\W}{\,\mathrm{W}}
\safemath{\J}{\,\mathrm{J}}
\safemath{\K}{\,\mathrm{K}}
\safemath{\bit}{\,\mathrm{bit}}

\safemath{\define}{\triangleq}                  

\safemath{\equivalent}{\sim}
\safemath{\distas}{\sim}                                        

\safemath{\reals}{\mathbb{R}}
\safemath{\positivereals}{\mathbb{R}^{+}}
\safemath{\integers}{\mathbb{Z}}
\safemath{\posint}{\mathbb{Z}_{+}}
\safemath{\naturals}{\mathbb{N}}
\safemath{\complexset}{\mathbb{C}}

\safemath{\setA}{\mathcal{A}}
\safemath{\setB}{\mathcal{B}}
\safemath{\setC}{\mathcal{C}}
\safemath{\setD}{\mathcal{D}}
\safemath{\setE}{\mathcal{E}}
\safemath{\setF}{\mathcal{F}}
\safemath{\setG}{\mathcal{G}}
\safemath{\setH}{\mathcal{H}}
\safemath{\setI}{\mathcal{I}}
\safemath{\setJ}{\mathcal{J}}
\safemath{\setK}{\mathcal{K}}
\safemath{\setL}{\mathcal{L}}
\safemath{\setM}{\mathcal{M}}
\safemath{\setN}{\mathcal{N}}
\safemath{\setO}{\mathcal{O}}
\safemath{\setP}{\mathcal{P}}
\safemath{\setQ}{\mathcal{Q}}
\safemath{\setR}{\mathcal{R}}
\safemath{\setS}{\mathcal{S}}
\safemath{\setT}{\mathcal{T}}
\safemath{\setU}{\mathcal{U}}
\safemath{\setV}{\mathcal{V}}
\safemath{\setW}{\mathcal{W}}
\safemath{\setX}{\mathcal{X}}
\safemath{\setY}{\mathcal{Y}}
\safemath{\setZ}{\mathcal{Z}}
\safemath{\emptySet}{\varnothing}

\safemath{\bma}{\mathbf{a}}
\safemath{\bmb}{\mathbf{b}}
\safemath{\bmc}{\mathbf{c}}
\safemath{\bmd}{\mathbf{d}}
\safemath{\bme}{\mathbf{e}}
\safemath{\bmf}{\mathbf{f}}
\safemath{\bmg}{\mathbf{g}}
\safemath{\bmh}{\mathbf{h}}
\safemath{\bmi}{\mathbf{i}}
\safemath{\bmj}{\mathbf{j}}
\safemath{\bmk}{\mathbf{k}}
\safemath{\bml}{\mathbf{l}}
\safemath{\bmm}{\mathbf{m}}
\safemath{\bmn}{\mathbf{n}}
\safemath{\bmo}{\mathbf{o}}
\safemath{\bmp}{\mathbf{p}}
\safemath{\bmq}{\mathbf{q}}
\safemath{\bmr}{\mathbf{r}}
\safemath{\bms}{\mathbf{s}}
\safemath{\bmt}{\mathbf{t}}
\safemath{\bmu}{\mathbf{u}}
\safemath{\bmv}{\mathbf{v}}
\safemath{\bmw}{\mathbf{w}}
\safemath{\bmx}{\mathbf{x}}
\safemath{\bmy}{\mathbf{y}}
\safemath{\bmz}{\mathbf{z}}

\bmdefine{\biad}{a}
\bmdefine{\bibd}{b}
\bmdefine{\bicd}{c}
\bmdefine{\bidd}{d}
\bmdefine{\bied}{e}
\bmdefine{\bifd}{f}
\bmdefine{\bigd}{g}
\bmdefine{\bihd}{h}
\bmdefine{\biid}{i}
\bmdefine{\bijd}{j}
\bmdefine{\bikd}{k}
\bmdefine{\bild}{l}
\bmdefine{\bimd}{m}
\bmdefine{\bind}{n}
\bmdefine{\biod}{o}
\bmdefine{\bipd}{p}
\bmdefine{\biqd}{q}
\bmdefine{\bird}{r}
\bmdefine{\bisd}{s}
\bmdefine{\bitd}{t}
\bmdefine{\biud}{u}
\bmdefine{\bivd}{v}
\bmdefine{\biwd}{w}
\bmdefine{\bixd}{x}
\bmdefine{\biyd}{y}
\bmdefine{\bizd}{z}

\bmdefine{\bixid}{\xi}
\bmdefine{\bilambdad}{\lambda}
\bmdefine{\bimud}{\mu}
\bmdefine{\bithetad}{\theta}
\bmdefine{\biphid}{\phi}

\safemath{\bmia}{\biad}
\safemath{\bmib}{\bibd}
\safemath{\bmic}{\bicd}
\safemath{\bmid}{\bidd}
\safemath{\bmie}{\bied}
\safemath{\bmif}{\bifd}
\safemath{\bmig}{\bigd}
\safemath{\bmih}{\bihd}
\safemath{\bmii}{\biid}
\safemath{\bmij}{\bijd}
\safemath{\bmik}{\bikd}
\safemath{\bmil}{\bild}
\safemath{\bmim}{\bimd}
\safemath{\bmin}{\bind}
\safemath{\bmio}{\biod}
\safemath{\bmip}{\bipd}
\safemath{\bmiq}{\biqd}
\safemath{\bmir}{\bird}
\safemath{\bmis}{\bisd}
\safemath{\bmit}{\bitd}
\safemath{\bmiu}{\biud}
\safemath{\bmiv}{\bivd}
\safemath{\bmiw}{\biwd}
\safemath{\bmix}{\bixd}
\safemath{\bmiy}{\biyd}
\safemath{\bmiz}{\bizd}

\safemath{\bmxi}{\bixid}
\safemath{\bmlambda}{\bilambdad}
\safemath{\bmmu}{\bimud}
\safemath{\bmtheta}{\bithetad}
\safemath{\bmphi}{\biphid}

\safemath{\bA}{\mathbf{A}}
\safemath{\bB}{\mathbf{B}}
\safemath{\bC}{\mathbf{C}}
\safemath{\bD}{\mathbf{D}}
\safemath{\bE}{\mathbf{E}}
\safemath{\bF}{\mathbf{F}}
\safemath{\bG}{\mathbf{G}}
\safemath{\bH}{\mathbf{H}}
\safemath{\bI}{\mathbf{I}}
\safemath{\bJ}{\mathbf{J}}
\safemath{\bK}{\mathbf{K}}
\safemath{\bL}{\mathbf{L}}
\safemath{\bM}{\mathbf{M}}
\safemath{\bN}{\mathbf{N}}
\safemath{\bO}{\mathbf{O}}
\safemath{\bP}{\mathbf{P}}
\safemath{\bQ}{\mathbf{Q}}
\safemath{\bR}{\mathbf{R}}
\safemath{\bS}{\mathbf{S}}
\safemath{\bT}{\mathbf{T}}
\safemath{\bU}{\mathbf{U}}
\safemath{\bV}{\mathbf{V}}
\safemath{\bW}{\mathbf{W}}
\safemath{\bX}{\mathbf{X}}
\safemath{\bY}{\mathbf{Y}}
\safemath{\bZ}{\mathbf{Z}}

\safemath{\bZero}{\mathbf{0}}

\bmdefine{\biAd}{A}
\bmdefine{\biBd}{B}
\bmdefine{\biCd}{C}
\bmdefine{\biDd}{D}
\bmdefine{\biEd}{E}
\bmdefine{\biFd}{F}
\bmdefine{\biGd}{G}
\bmdefine{\biHd}{H}
\bmdefine{\biId}{I}
\bmdefine{\biJd}{J}
\bmdefine{\biKd}{K}
\bmdefine{\biLd}{L}
\bmdefine{\biMd}{M}
\bmdefine{\biOd}{N}
\bmdefine{\biPd}{O}
\bmdefine{\biQd}{P}
\bmdefine{\biRd}{R}
\bmdefine{\biSd}{S}
\bmdefine{\biTd}{T}
\bmdefine{\biUd}{U}
\bmdefine{\biVd}{V}
\bmdefine{\biWd}{W}
\bmdefine{\biXd}{X}
\bmdefine{\biYd}{Y}
\bmdefine{\biZd}{Z}

\bmdefine{\biDelta}{\Delta}
\bmdefine{\biLambda}{\Lambda}
\bmdefine{\biPhi}{\Phi}
\bmdefine{\biSigma}{\Sigma}
\bmdefine{\biOmega}{\Omega}
\bmdefine{\biTheta}{\Theta}

\safemath{\bimA}{\biAd}
\safemath{\bimB}{\biBd}
\safemath{\bimC}{\biCd}
\safemath{\bimD}{\biDd}
\safemath{\bimE}{\biEd}
\safemath{\bimF}{\biFd}
\safemath{\bimG}{\biGd}
\safemath{\bimH}{\biHd}
\safemath{\bimI}{\biId}
\safemath{\bimJ}{\biJd}
\safemath{\bimK}{\biKd}
\safemath{\bimL}{\biLd}
\safemath{\bimM}{\biMd}
\safemath{\bimN}{\biNd}
\safemath{\bimO}{\biOd}
\safemath{\bimP}{\biPd}
\safemath{\bimQ}{\biQd}
\safemath{\bimR}{\biRd}
\safemath{\bimS}{\biSd}
\safemath{\bimT}{\biTd}
\safemath{\bimU}{\biUd}
\safemath{\bimV}{\biVd}
\safemath{\bimW}{\biWd}
\safemath{\bimX}{\biXd}
\safemath{\bimY}{\biYd}
\safemath{\bimZ}{\biZd}

\safemath{\bDelta}{\bielta}
\safemath{\bLambda}{\biLambda}
\safemath{\bPhi}{\biPhi}
\safemath{\bSigma}{\biSigma}
\safemath{\bOmega}{\biOmega}
\safemath{\bTheta}{\biTheta}

\safemath{\veca}{\bma}
\safemath{\vecb}{\bmb}
\safemath{\vecc}{\bmc}
\safemath{\vecd}{\bmd}
\safemath{\vece}{\bme}
\safemath{\vecf}{\bmf}
\safemath{\vecg}{\bmg}
\safemath{\vech}{\bmh}
\safemath{\veci}{\bmi}
\safemath{\vecj}{\bmj}
\safemath{\veck}{\bmk}
\safemath{\vecl}{\bml}
\safemath{\vecm}{\bmm}
\safemath{\vecn}{\bmn}
\safemath{\veco}{\bmo}
\safemath{\vecp}{\bmp}
\safemath{\vecq}{\bmq}
\safemath{\vecr}{\bmr}
\safemath{\vecs}{\bms}
\safemath{\vect}{\bmt}
\safemath{\vecu}{\bmu}
\safemath{\vecv}{\bmv}
\safemath{\vecw}{\bmw}
\safemath{\vecx}{\bmx}
\safemath{\vecy}{\bmy}
\safemath{\vecz}{\bmz}
\safemath{\vecZero}{\bZero}

\safemath{\vecxi}{\bmxi}
\safemath{\veclambda}{\bmlambda}
\safemath{\vecmu}{\bmmu}
\safemath{\vectheta}{\bmtheta}
\safemath{\vecphi}{\bmphi}

\safemath{\matA}{\bA}
\safemath{\matB}{\bB}
\safemath{\matC}{\bC}
\safemath{\matD}{\bD}
\safemath{\matE}{\bE}
\safemath{\matF}{\bF}
\safemath{\matG}{\bG}
\safemath{\matH}{\bH}
\safemath{\matI}{\bI}
\safemath{\matJ}{\bJ}
\safemath{\matK}{\bK}
\safemath{\matL}{\bL}
\safemath{\matM}{\bM}
\safemath{\matN}{\bN}
\safemath{\matO}{\bO}
\safemath{\matP}{\bP}
\safemath{\matQ}{\bQ}
\safemath{\matR}{\bR}
\safemath{\matS}{\bS}
\safemath{\matT}{\bT}
\safemath{\matU}{\bU}
\safemath{\matV}{\bV}
\safemath{\matW}{\bW}
\safemath{\matX}{\bX}
\safemath{\matY}{\bY}
\safemath{\matZ}{\bZ}
\safemath{\matZero}{\bZero}

\safemath{\matDelta}{\bDelta}
\safemath{\matLambda}{\bLambda}
\safemath{\matPhi}{\bPhi}
\safemath{\matSigma}{\bSigma}
\safemath{\matOmega}{\bOmega}
\safemath{\matTheta}{\bTheta}

\safemath{\matIdentity}{\matI}

\newcommand{\utility[2]}{u_{#1}(#2)}
\newcommand{\utilityStar[1]}{u^*_{#1}}

\newcommand{\rate[1]}{R_{#1}}
\safemath{\infobits}{D}
\safemath{\totalbits}{M}
\newcommand{\power[1]}{p_{#1}}
\newcommand{\powerStar[1]}{p^*_{#1}}
\newcommand{\SINR[1]}{\gamma_{#1}}
\newcommand{\SINRStar[1]}{\gamma^*_{#1}}
\safemath{\SINRStarInf}{\overline{\SINR[]}^*}
\newcommand{\efficiencyFunction[1]}{f\left({#1}\right)}
\newcommand{\efficiencyFunctionPrime[1]}{f^\prime(#1)}

\newcommand{\channelresponse[2]}{c_{#1}(#2)}
\safemath{\chiptime}{T_c}
\safemath{\srake}{\pathno_S}
\newcommand{\delay[1]}{\tau_{#1}}

\safemath{\SP}{\text{SP}}
\safemath{\SI}{\text{SI}}
\safemath{\MAI}{\text{MAI}}

\newcommand{\hSP[1]}{h_{#1}^{(\SP)}}
\newcommand{\hSI[1]}{h_{#1}^{(\SI)}}
\newcommand{\hMAI[1]}{h_{#1}^{(\MAI)}}
\safemath{\varnoise}{\sigma^2}
\newcommand{\vectornorm}[1]{\left|\left|{#1}\right|\right|}
\newcommand{\vecpathgain[1]}{\bmalpha_{#1}}
\newcommand{\vecrakecoeff[1]}{\bmbeta_{#1}}
\newcommand{\matpathgain[1]}{\matA_{#1}}
\newcommand{\matrakecoeff[1]}{\matB_{#1}}
\newcommand{\matCoeffHsi}{\matPhi}
\newcommand{\coeffHsi[1]}{\phi_{#1}}

\safemath{\game}{G}
\safemath{\userset}{\setK}
\newcommand{\pmin[1]}{\underline{p}_{#1}}
\newcommand{\pmax[1]}{\overline{p}_{#1}}
\newcommand{\powerset[1]}{P_{#1}}
\newcommand{\powervect[1]}{\vecp_{#1}}

\newcommand{\functionGamma[1]}{\Gamma\left({#1}\right)}
\safemath{\powerTimesHsp}{q}
\safemath{\varq}{\sigma^2_\powerTimesHsp}
\safemath{\meanq}{\eta_\powerTimesHsp}

\safemath{\Po}{P_o}

\newcommand{\functionMu[1]}{\mu\left({#1}\right)}
\newcommand{\functionNu[1]}{\nu\left({#1}\right)}
\newcommand{\functionNuO[1]}{\nu_0\left({#1}\right)}

\newcommand{\varUser[1]}{\sigma^2_{{#1}}}

\safemath{\PDPratio}{\Lambda}
\newcommand{\pathgain[2]}{\alpha_{#1}^{(#2)}}
\newcommand{\rakecoeff[2]}{\beta_{#1}^{(#2)}}
\safemath{\pathno}{L}
\safemath{\prake}{\pathno_P}
\safemath{\Pratio}{r}
\safemath{\userno}{K}
\safemath{\frameno}{N_f}
\safemath{\pulseno}{N_c}
\safemath{\frametime}{T_f}
\safemath{\gain}{N}
\safemath{\processingMatrix}{\bG}

\safemath{\loadFactor}{\rho}

\newcommand{\SIratio[1]}{\gamma_{0,{#1}}}

\safemath{\as}{\stackrel{a.s.}{\rightarrow}}
\safemath{\loss}{\Phi}
\safemath{\linearLoss}{\varphi}
\newcommand{\distance[1]}{d_{#1}}

\newcommand{\txSignal[1]}{s_{tx}^{({#1})}(t)}
\newcommand{\txPulse[1]}{w_{tx}({#1})}
\newcommand{\polarityVector[1]}{\vecd_{#1}}
\newcommand{\polarityChip[2]}{d_{#2}^{(#1)}}
\newcommand{\thVector[1]}{\vecc_{#1}}
\newcommand{\thChip[2]}{c_{#2}^{({#1})}}
\newcommand{\infoBit[2]}{b_{#2}^{({#1})}}

\newcommand{\spreadingChip[2]}{s_{#2}^{({#1})}}

\newcounter{mytempeqncnt}

\begin{document}

\title{Performance Comparison of Energy-Efficient Power Control
for CDMA and Multiuser UWB Networks\thanks{This research
  was supported in part by the
  U. S. Air Force Research Laboratory under Cooperative
  Agreement No. FA8750-06-1-0252, and in part by the
  Network of Excellence in Wireless Communications NEWCOM
  of the European Commission FP6 under Contract No. 507325.}}

\author{
  \authorblockN{Giacomo Bacci,\authorrefmark{1} 
    Marco Luise\authorrefmark{1} and H.~Vincent Poor\authorrefmark{2}}
  \authorblockA{\authorrefmark{1} University of Pisa - 
    Dip. Ingegneria dell'Informazione - Via Caruso - 56122 Pisa, Italy\\
    Email: giacomo.bacci@iet.unipi.it; marco.luise@iet.unipi.it}
  \authorblockA{\authorrefmark{2} Princeton University - 
    Dept. of Electrical Engineering - Olden Street - 08544 Princeton, NJ, USA\\
    Email: poor@princeton.edu}
}

\maketitle

\begin{abstract}
This paper studies the performance of a wireless data network using 
energy-efficient power control techniques when different multiple access 
schemes, namely direct-sequence code division multiple access (DS-CDMA) and
impulse-radio ultrawideband (IR-UWB), are considered. Due to the large 
bandwidth of the system, the multipath channel is assumed to be 
frequency-selective. By making use of noncooperative game-theoretic models and
large-system analysis tools, explicit expressions for the achieved utilities
at the Nash equilibrium are derived in terms of the network parameters. 
A measure of the loss of DS-CDMA with respect to IR-UWB is proposed, 
which proves substantial equivalence between the two schemes. 
Simulation results are provided to validate the analysis.
\end{abstract}

\section{Introduction}\label{sec:intro}

The increasing demand for high-speed data services in wireless networks
calls for multiple access schemes with efficient resource allocation and 
interference mitigation. Both direct-sequence code division multiple access
(DS-CDMA) and impulse-radio ultrawideband (IR-UWB) are considered to be
potential candidates for such next-generation high-speed networks. 
Design of reliable systems must include transmitter power
control, which aims to allow each mobile terminal to achieve the required
quality of service at the uplink receiver while minimizing power consumption. 
Scalable techniques for energy-efficient power control can be derived 
using game theory \cite{mackenzie, saraydar2, bacci1}.

This paper compares the performance of game-theoretic power control schemes in
the uplink of an infrastructure network using either DS-CDMA or IR-UWB as a
multiple access technique. The performance index here is represented by the
achieved utility at the Nash equilibrium, where utility is defined as the
ratio of the throughput to the transmit power. Due to the large bandwidth 
occupancy \cite{molisch}, the channel fading is assumed to be 
frequency-selective. Resorting to a large-system analysis \cite{bacci1}, systems 
with equal spreading factor operating in a dense multipath environment are 
compared. Both analytical and numerical results show that, even though 
UWB-based networks always outperform CDMA-based systems, the difference 
between achieved utilities is so slight that equivalence in terms of energy
efficiency can be assumed.

The remainder of the paper is organized as follows. The system model is 
described in Sect. \ref{sec:model}. Sect. \ref{sec:npcg} contains the main
results of the proposed noncooperative power control game. A comparison 
between the energy efficiency of the two considered multiple access schemes
is performed in Sect. \ref{sec:comparison}, where also simulation results 
are shown. Some conclusions are drawn in Sect. \ref{sec:conclusion}.

\section{System Model}\label{sec:model}

\subsection{IR-UWB Wireless Networks}

We consider the uplink of a binary phase shift keying (BPSK) random 
time-hopping (TH) IR-UWB system with \userno users 
transmitting to a common concentration point. The transmitted signal 
from user $k$ is \cite{gezici1}
\be\label{eq:txSignal}
  \txSignal[k] = \sqrt{\frac{\power[k]\frametime}{\gain}}
  \sum_{n=-\infty}^{+\infty}{\polarityChip[k]{n}
    \infoBit[k]{\lfloor n/\frameno \rfloor}
  \txPulse[{t-n\frametime-\thChip[k]{n}\chiptime}]},
\ee
where $\txPulse[t]$ is the transmitted UWB pulse with duration \chiptime and
unit energy; $\power[k]$ is the transmit power of user $k$; \frametime is the 
frame time; $\infoBit[k]{\lfloor n/\frameno \rfloor}\in\{-1,+1\}$ is the 
information symbol transmitted by user $k$; and 
$\gain=\frameno\cdot\pulseno$ is the processing gain, where \frameno is the 
number of pulses representing one information symbol, and 
$\pulseno=\frametime/\chiptime$ denotes the number of possible pulse 
positions in a frame. Throughout this analysis, a system with polarity code 
randomization is considered \cite{nakache}. In particular, the polarity code 
for user $k$ is $\polarityVector[k]=
\{\polarityChip[k]{0},\cdots,\polarityChip[k]{\frameno-1}\}$, where the
$\polarityChip[k]{j}$'s are independent random variables taking values $\pm1$ 
with probability $1/2$. To allow the channel to be shared by many users 
without causing catastrophic collisions, a TH sequence 
$\thVector[k]=\{\thChip[k]{1},\cdots,\thChip[k]{\frameno}\}$ is assigned to 
each user, where $\thChip[k]{n}\in\{0,1,\cdots,\pulseno-1\}$ with equal 
probability $1/\pulseno$.

Defining a sequence $\{\spreadingChip[k]{n}\}$ as
\be\label{eq:spreadingChipsUWB}
  \spreadingChip[k]{n}=
  \begin{cases}
    \polarityChip[k]{\lfloor n/\pulseno \rfloor},&
    \thChip[k]{\lfloor n/\pulseno \rfloor\cdot\pulseno}=
    n-\lfloor n/\pulseno \rfloor\cdot\pulseno,\\
    0,&\text{otherwise},
  \end{cases}
\ee
we can express (\ref{eq:txSignal}) as
\be\label{eq:txSignal2}
  \txSignal[k] = \sqrt{\frac{\power[k]\frametime}{\gain}}
  \sum_{n=-\infty}^{+\infty}{\spreadingChip[k]{n}
    \infoBit[k]{\lfloor n/\gain \rfloor}
    \txPulse[{t-n\chiptime}]}.
\ee
It is worth noting that this system makes use of a ternary sequence 
$\{-1,0,+1\}$, where also the elements are dependent, due to the TH 
sequence.

The transmission is assumed to be over \emph{frequency-selective channels}, 
with the channel for user $k$ modeled as a tapped delay line:
\be
  \label{eq:channel}
  \channelresponse[k]{t} =
  \sum_{l=1}^{\pathno}{\pathgain[l]{k}\delta(t-(l-1)\chiptime-\delay[k])},
\ee
where \pathno is the number of channel paths; 
$\vecpathgain[k]=\tp{[\pathgain[1]{k},\dots,\pathgain[\pathno]{k}]}$ and
$\delay[k]$ are the fading coefficients and the delay of user $k$,
respectively. Considering a chip-synchronous scenario, symbols are
misaligned by an integer multiple of \chiptime: 
$\delay[k] = \Delta_k\chiptime$, for every $k$, where $\Delta_k$ is uniformly 
distributed in $\{0,1,\dots,\gain-1\}$. We also assume that 
channel characteristics remain unchanged over several symbol 
intervals \cite{gezici1}.

Due to the high resolution of UWB signals, multipath channels can have hundreds
of multipath components, especially in indoor environments. To mitigate the 
effect of multipath fading as much as possible, we consider an access point 
where \userno Rake receivers\cite{proakis} are used.\footnote{For ease of 
calculation, perfect channel estimation is considered throughout the paper.} 
The Rake receiver for user $k$ is in general composed of \pathno coefficients, 
where the vector $\vecrakecoeff[k]=\processingMatrix\cdot\vecpathgain[k]=
\tp{[\rakecoeff[1]{k},\dots,\rakecoeff[\pathno]{k}]}$ represents the combining 
weights for user $k$, and the $\pathno\times\pathno$ matrix 
\processingMatrix depends on the type of Rake receiver employed. 

The signal-to-interference-plus-noise ratio (SINR) of the $k$th user 
at the output of the Rake receiver can be well 
approximated (for large \frameno, typically at least 5) by \cite{gezici1}
\be
  \label{eq:sinr}
  \SINR[k] = \frac{\hSP[k]\power[k]}{\displaystyle{\hSI[k]\power[k] +
      \sum_{\substack{j=1\\j\neq k}}^{\userno}{\hMAI[kj]\power[j]} +
      \sigma^2}},
\ee
where \varnoise is the variance of the additive white Gaussian
noise (AWGN); and the terms due to signal part (SP),
self-interference (SI), and multiple access interference (MAI), are
\begin{align}
  \label{eq:hSP}
  \hSP[k] &= \herm{\vecrakecoeff[k]}\cdot\vecpathgain[k],\\
  \label{eq:hSI}
  \hSI[k] &= \frac{1}{\gain}
  \frac{\vectornorm{\matCoeffHsi\cdot
      \left(\herm{\matrakecoeff[k]}\cdot\vecpathgain[k]+
      \herm{\matpathgain[k]}\cdot\vecrakecoeff[k]\right)}^2}
       {\herm{\vecrakecoeff[k]}\cdot\vecpathgain[k]},\\
  \label{eq:hMAI}
  \hMAI[kj] &= \frac{1}{\gain}
  \frac{\vectornorm{\herm{\matrakecoeff[k]}\cdot\vecpathgain[j]}^2
  + \vectornorm{\herm{\matpathgain[j]}\cdot\vecrakecoeff[k]}^2
  + \left|\herm{\vecrakecoeff[k]}\cdot\vecpathgain[j]\right|^2}
  {\herm{\vecrakecoeff[k]}\cdot\vecpathgain[k]},
\end{align}
respectively, where the matrices
\begin{align}
  \label{eq:matrixA}
  \matpathgain[k] &=
  \begin{pmatrix}
    \pathgain[\pathno]{k}&\cdots&\cdots&\pathgain[2]{k}\\
    0&\pathgain[\pathno]{k}&\cdots&\pathgain[3]{k}\\
    \vdots&\ddots&\ddots&\vdots\\
    0&\cdots&0&\pathgain[\pathno]{k}\\
    0&\cdots&\cdots&0
  \end{pmatrix},\\
  \label{eq:matrixB}
  \matrakecoeff[k] &=
  \begin{pmatrix}
    \rakecoeff[\pathno]{k}&\cdots&\cdots&\rakecoeff[2]{k}\\
    0&\rakecoeff[\pathno]{k}&\cdots&\rakecoeff[3]{k}\\
    \vdots&\ddots&\ddots&\vdots\\
    0&\cdots&0&\rakecoeff[\pathno]{k}\\
    0&\cdots&\cdots&0
  \end{pmatrix},\\
  \label{eq:matrixPhi}
  \matCoeffHsi &=
  \diag\left\{\coeffHsi[1],\dots,\coeffHsi[\pathno-1]\right\}\text{, and}
  \quad \coeffHsi[l]=\sqrt{\frac{\min\{\pathno-l,\pulseno\}}{\pulseno}},
\end{align}
have been introduced for convenience of notation.

\subsection{DS-CDMA Wireless networks}

In order to perform a fair comparison, the uplink of a random DS-CDMA system 
with spreading factor \gain and \userno users is considered.
It can be noticed that (\ref{eq:txSignal2}) can represent a DS-CDMA system 
with processing gain \gain by considering the special case when 
$\frametime=\chiptime$ (and thus $\pulseno=1$) \cite{gezici1}. As is apparent 
from (\ref{eq:spreadingChipsUWB}), using $\pulseno=1$ yields the elements of
$\{\spreadingChip[k]{n}\}$ to be binary independent and identical 
distributed (i.i.d.). 

Hence, in a dense frequency-selective multipath environment, the SINR of
user $k$ at the output of the Rake receiver is also represented by 
(\ref{eq:sinr}), under the conditions $\pulseno=1$, $\gain=\frameno$.

\section{The Noncooperative Power Control Game}\label{sec:npcg}

Consider now the application of noncooperative power control techniques to the 
wireless networks described above. Focusing on mobile terminals, where it is 
often more important to maximize the number of bits transmitted per Joule of 
energy consumed than to maximize throughput, a game-theoretic energy-efficient 
approach like the one described in \cite{bacci1} can be considered.

\begin{figure*}[!t]
  \normalsize
  \setcounter{mytempeqncnt}{\value{equation}}
  \setcounter{equation}{18}
  \begin{subnumcases}{\!\!\!\!\!\!\!
      \functionNu[{\PDPratio,\Pratio,\loadFactor}]=}
    \label{eq:nu1st}
      \tfrac{
        \PDPratio\left(\PDPratio^{\loadFactor}-1\right)
          \left(4\PDPratio^{2\Pratio}+3\PDPratio^{\loadFactor}-1\right)
          -2\PDPratio^{\Pratio+\loadFactor}\left(\PDPratio^{\Pratio}+
          3\PDPratio-1\right)
          \loadFactor\log{\PDPratio}}
           {2\left(\PDPratio^{\Pratio}-1\right)^2\loadFactor
             \PDPratio^{1+\loadFactor}\log{\PDPratio}},
           &\!\!\!\!\!\text{if $0\le\loadFactor\le\min(\Pratio,1-\Pratio)$};\\
    \label{eq:nu2nd}
      \tfrac{
        \PDPratio\left(4\PDPratio^{\loadFactor}-1\right)
          \left(\PDPratio^{2\Pratio}-1\right)
          -2\PDPratio^{\Pratio+\loadFactor}
          \left(3\PDPratio\Pratio-\loadFactor+\PDPratio^{\Pratio}
          \loadFactor\right)
          \log{\PDPratio}}
           {2\left(\PDPratio^{\Pratio}-1\right)^2\loadFactor
             \PDPratio^{1+\loadFactor}\log{\PDPratio}},
           &\!\!\!\!\!\text{if $\Pratio\le\loadFactor\le1-\Pratio$ 
             and}\nonumber\\
           &\!\!\!\!\!\,\,\,\,\,\,\text{$\Pratio\le1/2$};\\
    \label{eq:nu3rd}
      \tfrac{
        -4\PDPratio^{2+2\Pratio}-4\PDPratio^{2+\loadFactor}+
          \PDPratio^{2(\Pratio+\loadFactor)}+4
          \PDPratio^{2+2\Pratio+\loadFactor}+
          3\PDPratio^{2+2\loadFactor}-2\PDPratio^{1+\Pratio+\loadFactor}
          \left(\Pratio+3\PDPratio\loadFactor+\PDPratio^{\Pratio}
          \loadFactor-1\right)
          \log{\PDPratio}}
           {2\left(\PDPratio^{\Pratio}-1\right)^2\loadFactor
             \PDPratio^{2+\loadFactor}\log{\PDPratio}},
           &\!\!\!\!\!\text{if $1-\Pratio\le\loadFactor\le\Pratio$ 
             and}\nonumber\\
           &\!\!\!\!\!\,\,\,\,\,\,\text{$\Pratio\ge1/2$};\\
    \label{eq:nu4th}
      \tfrac{
        -\PDPratio^{2+2\Pratio}-4\PDPratio^{2+\loadFactor}+
          \PDPratio^{2(\Pratio+\loadFactor)}+
          4\PDPratio^{2+2\Pratio+\loadFactor}-
          2\PDPratio^{1+\Pratio+\loadFactor}
          \left(\Pratio+3\PDPratio\loadFactor+
          \PDPratio^{\Pratio}\loadFactor-1\right)
          \log{\PDPratio}}
           {2\left(\PDPratio^{\Pratio}-1\right)^2
             \loadFactor\PDPratio^{2+\loadFactor}\log{\PDPratio}}
           &\!\!\!\!\!\text{if $\max(\Pratio,1-\Pratio)\le\loadFactor\le1$};\\
    \label{eq:nu5th}
      \tfrac{2\PDPratio\left(\PDPratio^{2\Pratio}-1\right)-
        \left(\PDPratio^{\Pratio}+\Pratio+3\PDPratio\Pratio-1\right)
        \PDPratio^{\Pratio}\log{\PDPratio}}
           {\left(\PDPratio^{\Pratio}-1\right)^2
             \loadFactor\PDPratio\log{\PDPratio}},
           &\!\!\!\!\!\text{if $\loadFactor\ge1$}.
  \end{subnumcases}
  \setcounter{equation}{\value{mytempeqncnt}}
  \hrulefill
  \vspace*{4pt}
\end{figure*}

\subsection{Analysis of the Nash equilibrium}

It is possible to formulate a noncooperative power control game in which 
each user seeks to maximize its own utility function. Let 
$\game = [\userset, \{\powerset[k]\}, 
\{\utility[k]{\powervect[]}\}]$ be the proposed noncooperative game where 
$\userset=\{1,\dots,\userno\}$ is the index set for the users; 
$\powerset[k]=[\pmin[k], \pmax[k]]$ is the strategy set, with 
$\pmin[k]$ and $\pmax[k]$ denoting minimum and maximum power constraints, 
respectively; and $\utility[k]{\powervect[]}$ is the payoff function for 
user $k$ \cite{saraydar2}, defined as
\be
  \label{eq:utility}
  \utility[k]{\powervect[]}=\frac{\infobits}{\totalbits}\rate[k]
  \frac{\efficiencyFunction[{\SINR[k]}]}{\power[k]},
\ee
where $\powervect[]=[\power[1],\dots,\power[\userno]]$ is the vector of 
transmit powers; $\infobits$ and $\totalbits$ are the number of information 
bits and the total number of bits in a packet, respectively; $\rate[k]$ and 
$\SINR[k]$ are the transmission rate and the SINR for the $k$th user, 
respectively; and $\efficiencyFunction[{\SINR[k]}]$ is the efficiency function 
representing the packet success rate (PSR), i.e., the probability that a 
packet is received without an error. Throughout this analysis, we assume 
$\pmin[k]=0$ and $\pmax[k]=\pmax[]$ for all $k\in\userset$.

Provided that the efficiency function is increasing, S-shaped, and 
continuously differentiable, with $\efficiencyFunction[0]=0$, 
$\efficiencyFunction[+\infty]=1$, and $\efficiencyFunctionPrime[0]=
d\efficiencyFunction[{\SINR[k]}]/d\SINR[k]|_{\SINR[k]=0}=0$, 
the solution of the maximization problem 
$\max_{\power[k]\in\powerset[k]} \utility[k]{\powervect[]}$ for 
$k=1,\dots,\userno$ is \cite{bacci1}
\be
  \label{eq:powerStar}
  \powerStar[k]=\min\left\{
  \frac{\SINRStar[k]\left(\sum_{j\neq k}{\hMAI[kj]\power[j]}+
    \sigma^2\right)}{\hSP[k]\left(1-\SINRStar[k]/\SIratio[k]\right)}, 
  \pmax[]\right\},
\ee
where
\be\label{eq:SIratio}
  \SIratio[k]=\frac{\hSP[k]}{\hSI[k]}
  =\gain\cdot
  \frac{(\herm{\vecrakecoeff[k]}\cdot\vecpathgain[k])^2}
       {\vectornorm{\matCoeffHsi\cdot
           \left(\herm{\matrakecoeff[k]}\cdot\vecpathgain[k]+
           \herm{\matpathgain[k]}\cdot\vecrakecoeff[k]\right)}^2}
       \ge1
\ee
and $\SINRStar[k]$ is the solution of 
\be\label{eq:f_der}
  \efficiencyFunctionPrime[{\SINRStar[k]}]
  \SINRStar[k]\left(1-\SINRStar[k]/\SIratio[k]\right)=
  \efficiencyFunction[{\SINRStar[k]}],
\ee
where $\efficiencyFunctionPrime[{\SINRStar[k]}]=
d\efficiencyFunction[{\SINR[k]}]/d\SINR[k]|_{\SINR[k]=\SINRStar[k]}$.
Since $\SINRStar[k]$ depends only on $\SIratio[k]$, for convenience of 
notation a function $\functionGamma[\cdot]$ is defined such that 
$\SINRStar[k]=\functionGamma[{\SIratio[k]}]$ \cite{bacci1}.

\subsection{Large-System Analysis}

As can be verified in (\ref{eq:powerStar}), the amount of transmit power 
$\powerStar[k]$ required to 
achieve the target SINR $\SINRStar[k]$ will depend not only on 
$\hSP[k]$, but also on the SI term $\hSI[k]$ (through
$\SIratio[k]$) and the MAI (through $\hMAI[kj]$).
To derive quantitative results for the transmit powers independent of 
SI and MAI terms, it is possible to resort to the large-system analysis 
described in \cite{bacci1}.

For ease of calculation, the expressions derived in the remainder of the paper
consider the following assumptions:
\begin{itemize}
  \item The channel gains are assumed to be independent complex Gaussian 
    random variables with zero means and variances $\varUser[{k_l}]$, i.e., 
    $\pathgain[k]{l}\distas\circnorm(0, \varUser[{k_l}])$. This assumption 
    leads $|\pathgain[k]{l}|$ to be Rayleigh-distributed with 
    parameter $\varUser[{k_l}]/2$. Although channel modeling for wideband 
    systems is still an open issue, this hypothesis, appealing for its 
    analytical tractability, also provides a good approximation for multipath 
    propagation in UWB systems \cite{schuster}.

  \item The average power delay profile (aPDP) \cite{hashemi} is assumed to 
    decay exponentially, as is customary used in many UWB channel 
    models \cite{molisch}. This translates into the hypothesis 
    $\varUser[{k_l}]=\varUser[k]\cdot\PDPratio^{-\frac{l-1}{\pathno-1}}$,
    where $\PDPratio=\varUser[{k_1}]/\varUser[{k_\pathno}]$ and $\varUser[k]$
    depends on the distance between user $k$ and the access point.
  \item Partial-Rake (PRake) receivers with \prake fingers using maximal ratio 
    combining (MRC) are implemented at the access point. In other words, we 
    consider \processingMatrix to be a deterministic diagonal matrix, with
    \be\label{eq:prakeDef}
    \{\processingMatrix\}_{ll}=
    \begin{cases}
      1, & 1 \le l \le \Pratio\cdot\pathno,\\
      0, & \text{elsewhere},
    \end{cases}
    \ee
    where $\Pratio\triangleq\prake/\pathno$ and $0<\Pratio\le1$. It is worth 
    noting that, when $\Pratio=1$, an all-Rake (ARake) is implemented.
  \item As is typical in multiuser wideband systems, the number of users is 
    much smaller than the processing gain, i.e., $\gain\gg\userno$.
    This assumption can also be justified since the analysis is performed 
    for dense multipath environments, as shown in the following.
  \item The maximum transmit power $\pmax[]$ is assumed to be sufficiently 
    large.
\end{itemize}

Under the above hypotheses, a large-system analysis can be performed 
considering a dense multipath environment, with $\pathno\to\infty$. It 
turns out that the achieved utilities $\utilityStar[k]$ at the Nash 
equilibrium converge almost surely (a.s.) to \cite{bacci1, bacci2}
\begin{align}
  \label{eq:utilityLSA}
  \utilityStar[k]&\as\hSP[k]\cdot\frac{\infobits}{\totalbits}\rate[k]
  \frac{\efficiencyFunction[{\functionGamma[{\frac{\gain}
        {\functionNu[{\PDPratio,\Pratio,\loadFactor}]}}]}]}
       {\sigma^2\functionGamma[{\frac{\gain}
        {\functionNu[{\PDPratio,\Pratio,\loadFactor}]}}]}\nonumber\\
  &\qquad\times\!\left(\!1\!-\!\frac{\functionGamma[{\frac{\gain}
        {\functionNu[{\PDPratio,\Pratio,\loadFactor}]}}][(\userno\!-\!1)
    \functionMu[{\PDPratio,\Pratio}]\!+\!
    \functionNu[{\PDPratio,\Pratio,\loadFactor}]]}{\gain}\right)\!,
\end{align}
where $\Pratio\triangleq\prake/\pathno$, with $0\!<\!\Pratio\!\le\!1$;
$\loadFactor\triangleq\pulseno/\pathno$, with $0\!<\!\loadFactor\!<\!\infty$;
\be\label{eq:functionMu}
  \functionMu[{\PDPratio,\Pratio}]=
  \frac{\left(\PDPratio-1\right)\cdot\PDPratio^{\Pratio-1}}
       {\PDPratio^{\Pratio}-1};
\ee
and $\functionNu[{\PDPratio,\Pratio,\loadFactor}]$ is defined as in 
(\ref{eq:nu1st})-(\ref{eq:nu5th}), shown at the top of the page.

\section{Performance Comparison}\label{sec:comparison}

\subsection{Analytical Results}

The results derived in the previous section allow the performance of IR-UWB
and DS-CDMA systems to be compared in terms of achieved utilities at the
Nash equilibrium. 

For an IR-UWB system, the utility $\utilityStar[{k_{U}}]$ can be evaluated
using (\ref{eq:utilityLSA}). In the case of a DS-CDMA system, 
(\ref{eq:utilityLSA}) can still give the utility 
$\utilityStar[{k_{C}}]$, provided that 
$\functionNu[{\PDPratio,\Pratio,\loadFactor}]$ is replaced with
$\functionNuO[{\PDPratio,\Pratio}]$, defined as
\be\label{eq:functioNuO}
  \setcounter{equation}{20}
  \functionNuO[{\PDPratio,\Pratio}]=\lim_{\loadFactor\to0}
  \functionNu[{\PDPratio,\Pratio,\loadFactor}]
  =\frac{\PDPratio+\PDPratio^\Pratio-2\PDPratio^{1+\Pratio}}
             {\PDPratio-\PDPratio^{1+\Pratio}}.
\ee
This results is obtained letting \loadFactor go to $0$ in (\ref{eq:nu1st}).
The proof, omitted because of space limitation, can 
be derived using the theorems presented in \cite{bacci2} with $\pulseno=1$.

\begin{figure}
  \centering
  \includegraphics[width=9.0cm]{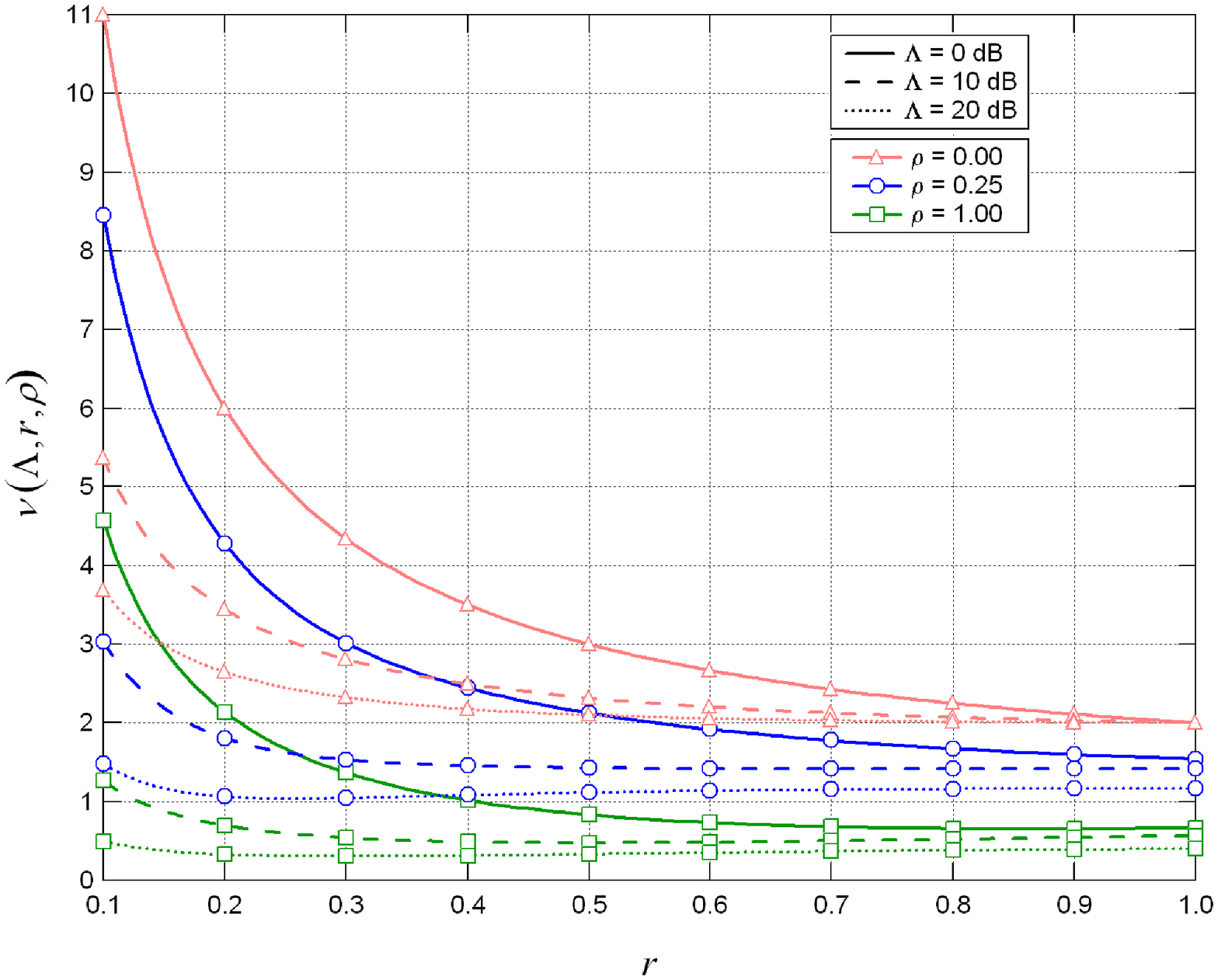}
  \caption{Shape of $\functionNu[{\PDPratio, \Pratio, \loadFactor}]$ 
    versus \Pratio for some values of \PDPratio and \loadFactor.}
  \label{fig:nu}
\end{figure}

Fig. \ref{fig:nu} shows the shape of 
$\functionNu[{\PDPratio, \Pratio, \loadFactor}]$ as a function of $\Pratio$ for
some values of $\PDPratio$ and $\loadFactor$. With a slight abuse of notation,
$\functionNuO[{\PDPratio,\Pratio}]$ is reported as 
$\functionNu[{\PDPratio,\Pratio,0}]$ (triangular markers), while
circles and square markers depict $\loadFactor=0.25$ and $\loadFactor=1.0$, 
respectively. As can be noted, $\functionNuO[{\PDPratio,\Pratio}]>
\functionNu[{\PDPratio,\Pratio,\loadFactor_1}]>
\functionNu[{\PDPratio,\Pratio,\loadFactor_2}]$ for any $\loadFactor_2>
\loadFactor_1>0$. This result is justified by the higher resistance to 
multipath due to increasing the length of a single 
frame \cite{gezici1, bacci1}. Furthermore, keeping \loadFactor fixed, 
$\functionNu[{\PDPratio,\Pratio,\loadFactor}]$ decreases both as
\PDPratio and as \Pratio increases. The first behavior makes sense,
since the effect of multipath (and thus of SI) is higher in channels
with lower \PDPratio. The second behavior reflects the fact that 
exploiting the diversity by adding a higher number of fingers (and thus 
increasing \Pratio) results in better mitigating the frequency-selective 
fading.

\begin{proposition}\label{prop:loss}
  When $\pathno\to\infty$, the loss \loss
  of a CDMA system with respect to (wrt) an IR-UWB scheme 
  with \pulseno possible pulse positions converges a.s. to
  \begin{align}
    \label{eq:loss}
    \loss&\triangleq
    10\log_{10}(\utilityStar[{k_{U}}]/\utilityStar[{k_{C}}])
    \as(10\log_{10}e)\cdot\linearLoss\qquad\text{[dB]}\\
    \intertext{where}
    \label{eq:linearLoss}
    \linearLoss&\triangleq\frac{\functionGamma[{\frac{\gain}
        {\functionNu[{\PDPratio,\Pratio,\loadFactor}]}}]\cdot
      \Delta\functionNu[{\PDPratio,\Pratio,\loadFactor}]}
    {\gain-\functionGamma[{\frac{\gain}
        {\functionNu[{\PDPratio,\Pratio,\loadFactor}]}}]\cdot
      [(\userno-1)
      \functionMu[{\PDPratio,\Pratio}]+
      \functionNu[{\PDPratio,\Pratio,\loadFactor}]]},
  \end{align}
  with $\Delta\functionNu[{\PDPratio,\Pratio,\loadFactor}]=
  \functionNuO[{\PDPratio,\Pratio}]-
  \functionNu[{\PDPratio,\Pratio,\loadFactor}]$.
\end{proposition}

\begin{proof}\label{pr:loss}
  Recalling (\ref{eq:f_der}), it can be noted that the slope of
  $\functionGamma[{\SIratio[k]}]$ is very small for large values of
  $\SIratio[k]$. Using the hypothesis $\gain\gg\userno>1$, a good approximation
  for $\functionGamma[{\gain/\functionNuO[{\PDPratio,\Pratio}]}]$ is
  $\functionGamma[{\gain/\functionNu[{\PDPratio,\Pratio,\loadFactor}]}]$.
  Therefore, using (\ref{eq:utilityLSA}),
  \begin{align}\label{eq:proof1}
    \frac{\utilityStar[{k_{U}}]}
    {\utilityStar[{k_{C}}]}&\approxeq
    \frac{\gain-\functionGamma[{\frac{\gain}
        {\functionNu[{\PDPratio,\Pratio,\loadFactor}]}}]\cdot
      [(\userno-1)
      \functionMu[{\PDPratio,\Pratio}]+
      \functionNu[{\PDPratio,\Pratio,\loadFactor}]]}
    {\gain-\functionGamma[{\frac{\gain}
        {\functionNu[{\PDPratio,\Pratio,\loadFactor}]}}]\cdot
      [(\userno-1)
      \functionMu[{\PDPratio,\Pratio}]+
      \functionNuO[{\PDPratio,\Pratio}]]}\\
    &=\frac{1}{1-\linearLoss},
  \end{align}
  with \linearLoss defined as in (\ref{eq:linearLoss}). Recalling that
  $\gain\gg1$, it is easy to verify that
  $\linearLoss \ll 1$. Hence, using a first-order Taylor series approximation, 
  the result (\ref{eq:loss}) is straightforward.
\end{proof}

As already specified (see also Fig. \ref{fig:nu}), 
$\Delta\functionNu[{\PDPratio,\Pratio,\loadFactor}]>0$ for any $\loadFactor>0$.
Proposition \ref{prop:loss} thus states that, using an equal spreading factor 
in the same multipath scenario, any UWB system outperforms the corresponding
CDMA schemes.

Nevertheless, typical values of the network parameters yield very small values
of \loss, especially as \gain increases.\footnote{As expected, larger spreading
factors better mitigate multipath effects.} Hence, using game-theoretic power 
control techniques, performance of the two multiple access schemes is 
practically equivalent.

The validity of these claims is verified in the next subsection using 
numerical simulations.

\begin{table}
  \renewcommand{\arraystretch}{1.3}
  \caption{List of parameters used in the simulations.}
  \label{tab:parameters}
  \centering
  \begin{tabular}{c|c}
    \hline
    $\totalbits$, total number of bits per packet & $100\,\text{b}$ \\
    \hline
    $\infobits$, number of information bits per packet & $100\,\text{b}$ \\
    \hline
    $\rate[k]=\rate[]$, bit rate & $100\,\text{kb/s}$ \\
    \hline
    $\sigma^2$, AWGN power at the receiver & 
    $5 \times 10^{-16}\,\text{W}$\\
    \hline
    $\pmax[]$, maximum power constraint & $1\,\mu\text{W}$ \\
    \hline
  \end{tabular}
\end{table}

\subsection{Numerical Results}

Simulations are performed using the iterative algorithm described in detail
in \cite{bacci1}. The systems we examine have the design parameters listed in
Table \ref{tab:parameters}. We use the efficiency function
$\efficiencyFunction[{\SINR[k]}]=(1-\text{e}^{-\SINR[k]/2})^\totalbits$
as a reasonable approximation to the PSR \cite{saraydar2, gezici1}. To model 
the UWB scenario, the channel gains are assumed as in Sect. \ref{sec:npcg}, 
with $\varUser[k]=0.3\distance[k]^{-2}$, where $\distance[k]$ is the distance 
between user $k$ and the access point. Distances are assumed to be uniformly 
distributed between $3$ and $30\,\text{m}$. 

\begin{figure}
  \centering
  \includegraphics[width=9.0cm]{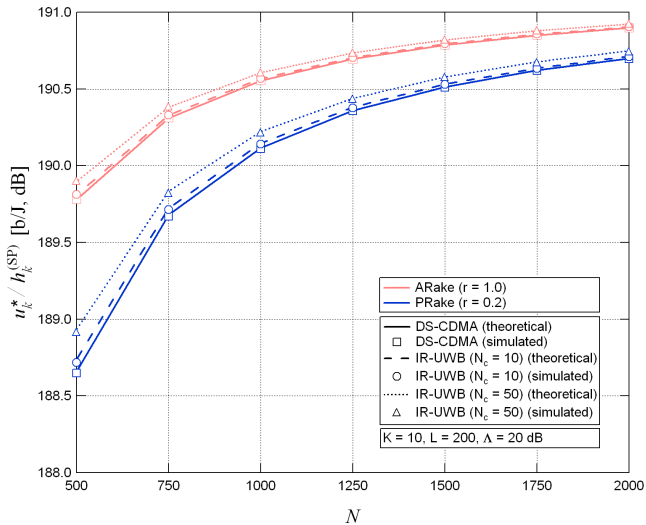}
  \caption{Comparison of normalized utilities versus processing gain
  for DS-CDMA and IR-UWB schemes.}
  \label{fig:utilityVsGain}
\end{figure}

Fig. \ref{fig:utilityVsGain} shows a comparison between analytical and 
simulated normalized utilities $\utilityStar[k]/\hSP[k]$ at the Nash 
equilibrium as a function of the spreading factor \gain. A network 
with $\userno=10$ users is considered, while the aPDP is assumed to 
be exponentially decaying with $\PDPratio=20\,\text{dB}$. The 
number of paths is $\pathno=200$, thus satisfying the large-system 
assumption. Red (light) and blue (dark) colors depict the cases 
ARake ($\Pratio=1$) and 
PRake ($\Pratio=0.2$), respectively. Lines represent theoretical results 
provided by (\ref{eq:utilityLSA}). In particular, solid lines show analytical 
values for DS-CDMA ($\pulseno=1$), while dashed and dotted lines report the
IR-UWB scenario, with $\pulseno=10$ and $\pulseno=50$, respectively. The 
markers show the simulation results averaged over $10\,000$ network 
realizations. It can be seen that the analytical results perfectly match
the actual performance of systems. As expected, the performance loss of 
DS-CDMA wrt IR-UWB is negligible (less than $1\,\text{dB}$) when
compared with the normalized achieved utilities. 
Furthermore, with \gain fixed, numerical results confirm
that a higher \Pratio provides smaller difference in performance between the
two multiple access schemes. Moreover, such loss decreases as \gain increases,
due to the inherent resistance to multipath, thus smoothing the 
performance behavior.

\begin{figure}
  \centering
  \includegraphics[width=9.0cm]{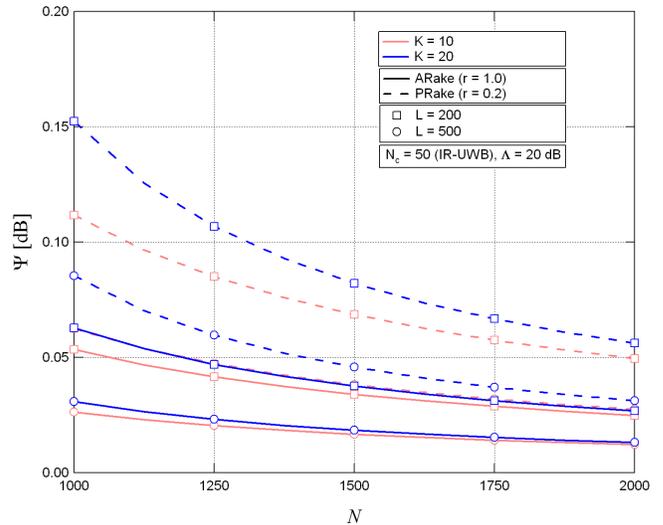}
  \caption{Performance loss of DS-CDMA wrt IR-UWB for 
    different values of the system parameters.}
  \label{fig:loss}
\end{figure}

Similar considerations can be made when observing the results shown in Fig.
\ref{fig:loss}. The loss of a DS-CDMA wrt an IR-UWB with 
$\pulseno=50$ is studied. The decay constant of the channel is assumed to
be $\PDPratio=20\,\text{dB}$. For the sake of presentation, only analytical 
results are reported. Red (light) and blue (dark) colors represent
$\userno=10$ and 
$\userno=20$, respectively. The solid lines depict the case ARake, while the
dashed lines show the case PRake ($\Pratio=0.2$). The square markers and the
circles report the results with $\pathno=200$ and $\pathno=500$ multiple paths,
respectively. It can be seen that the loss \loss is always very small. In 
addition, it is worth noting that, for both \gain and \pulseno fixed, \loss 
decreases as \pathno increases. This can be justified since IR-UWB access 
scheme cannot further mitigate the effects of denser and denser multipath 
in a \pulseno-fixed scenario, and thus its behavior is more similar to that of
DS-CDMA systems. This statement is not in contrast with the comments on Fig.
\ref{fig:utilityVsGain}, since the comparison would be completely different 
if we considered IR-UWB with fixed \gain and variable \pulseno. In fact,
if we choose \pulseno such that \loadFactor is constant accordingly to the 
increasing \pathno, \loss remains unchanged, as is apparent from 
(\ref{eq:loss}).

\section{Conclusion}\label{sec:conclusion}

In this paper, two multiple access schemes, namely DS-CDMA and IR-UWB, have 
been compared in the context of game-theoretic energy-efficient power control. 
We have used a large-system analysis to study the performance of a wireless 
data network using Rake receivers at the access point in a frequency-selective
fading channel. Considering systems with equal spreading factor, 
a measure of the loss of DS-CDMA scheme with respect to IR-UWB multiple
access technique has been derived which is dependent only on the network
parameters. Theoretical analysis, supported by experimental results,
shows that the considered multiple access schemes are practically equivalent
in terms of energy efficiency.

\vspace{0.1cm}

\begin{thebibliography}{99}


\bibitem{mackenzie}
A. B. MacKenzie and S. B. Wicker, ``Game theory in communications:
Motivation, explanation, and application to power control,'' in
\emph{Proc. IEEE Global Telecommun. Conf.}, 
San Antonio, TX, 2001, pp. 821-826.



\bibitem{saraydar2}
C. U. Saraydar, N. B. Mandayam and D. J. Goodman, ``Efficient power
control via pricing in wireless data networks,'' \emph{IEEE Trans. Commun.},
Vol. 50 (2), pp. 291-303, Feb. 2002.


\bibitem{bacci1}
G. Bacci, M. Luise, H. V. Poor and A. M. Tulino, ``Energy-efficient power 
control in impulse radio UWB wireless networks,'' preprint, Princeton 
University.
[Online]. Available: http://arxiv.org/pdf/cs/0701017.

\bibitem{molisch}
A. F. Molisch, J. R. Foerster and M. Pendergrass, ``Channel models for
ultrawideband personal area networks,'' \emph{IEEE Wireless Commun.}, Vol. 10
(6), pp. 14-21, Dec. 2003.

\bibitem{gezici1}
S. Gezici, H. Kobayashi, H. V. Poor and A. F. Molisch, ``Performance
evaluation of impulse radio UWB systems with pulse-based polarity
randomization,'' \emph{IEEE Trans. Signal Process.}, Vol. 53 (7), pp.
2537-2549, Jul. 2005.

\bibitem{nakache}
Y.-P. Nakache and A. F. Molisch, ``Spectral shape of UWB signals influence 
of modulation format, multiple access scheme, and pulse shape,'' in
\emph{Proc. IEEE Veh. Technol. Conf.}, Jeju, Korea, 2003, pp. 2510-2514.

\bibitem{proakis}
J. G. Proakis, \emph{Digital Communications}, 4th ed. \hskip 1em plus
0.5em minus 0.4em\relax New York, NY, USA: McGraw-Hill, 2001.

\bibitem{schuster}
U. G. Schuster and H. B\"{o}lcskei, ``Ultrawideband channel modeling
on the basis of information-theoretic criteria,'' \emph{IEEE Trans.
Wireless Commun.}, 2007, to appear. [Online]. Available:
http://www.nari.ee.ethz.ch/commth/pubs/p/schuster-tw06.

\bibitem{hashemi}
H. Hashemi, ``The indoor radio propagation channel,'' \emph{Proc. IEEE},
Vol. 81 (7), pp. 943-968, Jul. 1993.

\bibitem{bacci2}
G. Bacci, M. Luise and H. V. Poor, ``Performance of rake receivers in
IR-UWB networks using energy-efficient power control,'' preprint,
Princeton University. [Online]. Available: http://arxiv.org/pdf/cs/0701034.







\end{thebibliography}
\end{document}